\documentclass[pra,amssymb,amsmath,showpacs]{revtex4}

\usepackage{graphicx}
\usepackage{amssymb}
\usepackage{amsmath}
\usepackage{color}

\newcommand{\be}{\begin{equation}}
\newcommand{\ee}{\end{equation}}
\newcommand{\bea}{\begin{eqnarray}}
\newcommand{\eea}{\end{eqnarray}}

\begin{document}       

\title{Model of Low-pass Filtering of Local Field Potentials in Brain Tissue}

\author{C.~B\'edard$^{a}$, H.~Kr\"{o}ger$^{a}$$\footnote{Corresponding author, Email: hkroger@phy.ulaval.ca}$, A.~Destexhe$^{b}$} 

\affiliation{
$^{a}$ {\small\sl D\'{e}partement de Physique, Universit\'{e} Laval, Qu\'{e}bec, 
Qu\'{e}bec G1K 7P4, Canada} \\ 
$^{b}$ {\small\sl Unit\'e de Neurosciences Int\'egratives et Computationnelles, CNRS, \\
1 Avenue de la Terrasse, 91198 Gif-sur-Yvette, France} \\ \ \\
Revised version, \today
}

\begin{abstract}

Local field potentials (LFPs) are routinely measured experimentally in brain
tissue, and exhibit strong low-pass frequency filtering properties, with high
frequencies (such as action potentials) being visible only at very short
distances ($\approx$10~$\mu m$) from the recording electrode.  Understanding this
filtering is crucial to relate LFP signals with neuronal activity, but not much
is known about the exact mechanisms underlying this low-pass filtering.  In this
paper, we investigate a possible biophysical mechanism for the low-pass filtering
properties of LFPs.  We investigate the propagation of electric fields and its
frequency dependence close to the current source, i.e. at length scales in the
order of average interneuronal distance.  We take into account the presence of a
high density of cellular membranes around current sources, such as glial cells. 
By considering them as passive cells, we show that under the influence of the
electric source field, they respond by polarisation, i.e., creation of an induced
field.  Because of the finite velocity of ionic charge movement, this
polarization will not be instantaneous.  Consequently, the induced electric field
will be frequency-dependent, and much reduced for high frequencies.  Our model
establishes that with respect to frequency attenuation properties, this situation
is analogous to an equivalent RC-circuit, or better a system of coupled
RC-circuits.  We present a number of numerical simulations of induced electric
field for biologically realistic values of parameters, and show this frequency
filtering effect as well as the attenuation of extracellular potentials with
distance.  We suggest that induced electric fields in passive cells surrounding
neurons is the physical origin of frequency filtering properties of LFPs. 
Experimentally-testable predictions are provided allowing to verify the validity
of this model.

\end{abstract}


\maketitle 
      

\section{Introduction}
\label{sec:Intro}

Electric fields of the brain which are experimentally observable either on the
surface of the scalp, or by microelectrodes, are due to currents in dendrites and
the soma of cortical pyramidal cells \cite{Nunez81,Niedermeyer98}. Experimental
measurements of electric fields created in the brain distinguish three scenarios:
(i) Local field potentials (LFPs) denote the electric potential recorded locally
in the immediate neighborhood of neurons using microelectrodes of size comparable
to the cell body. (ii) The electrocorticogram (ECoG) refers to measurements of
the field using electrodes of a diameter of the size of about 1~mm, placed on the
cortical surface. (iii) The electroencephalogram (EEG) is measured at the surface
of the scalp using electrodes of a centimeter scale.  In the latter case, the
electric potential is recorded after conduction through cerobrospinal fluid,
cranium and scalp, and corresponds to the situation where the source of the
electric signals in the cortex is located far from the site of detection on the
scalp (at a scale $\lambda \approx 10^{3} d_{nn}$, where $d_{nn} = 0.027$ mm is
the average distance between cortical neurons).  

By contrast with the intracellular or membrane potential, which biophysical
properties have been extensively studied~\cite{Hille92,Johnston99,Koch99}, the
mechanisms underlying the genesis of LFPs are still unclear.  LFP recordings
routinely show strong high-frequency attenuation properties, because action
potentials are only visible for a few neurons immediately adjacent to the
electrode, while low-frequency components results from large populations of
neurons in the local network.  Using LFPs, it has been shown that low-frequency
oscillations (0-4~Hz) have a large scale coherence, while the coherence of higher
frequency (20-60~Hz) oscillations is short-ranged. In anesthetized animals, it
was shown that oscillations of up to 4~Hz have a coherence range in the order of
several millimeters, while oscillations of 20-60~Hz have a coherence of
sub-millimeter range \cite{Bullock89,Steriade96c}.  The local coherence of high
frequency oscillations was also shown in the visual system of anesthetized cats,
where gamma-oscillations (30-50~Hz) appear only within restricted cortical areas
and time windows \cite{Eckhorn88,Gray89}.  The same difference of coherence
between low and high frequency oscillations was also demonstrated in
non-anesthetized animals, respectively during sleep in a state of low-frequency
wave and waking \cite{Destexhe99}. Similar findings have been reported for human
EEG \cite{Achermann98}.  

A full understanding of the mechanisms underlying the genesis of EEG and LFP
signals is required to relate these signals with neuronal activity.  Several
models of EEG or LFP activity have been proposed previously (e.g., see
\cite{Nunez81,Nunez94,Rall68,Klee77,Lagerlund88,Destexhe98,Protopapas98}).  These
models always considered current sources embedded in a homogeneous extracellular
fluid.  In such homogeneous media, however, there cannot be any frequency
filtering property.  Extracellular space consists of a complex folding of
intermixed layers of fluids and membranes, while the extracellular fluid
represents only a few percent of the available
space~\cite{Braitenberg98,Peters91}.  Due to the complex nature of this medium,
it is very difficult to draw theories or to model LFPs properly, and one needs to
make approximations.  In a previous paper~\cite{Bedard04}, we considered current
sources with various continuous profiles of conductivity according to a spherical
symmetry, and showed that it can lead to low-pass frequency filtering.  This
showed that strong inhomogeneities in conductivity and permittivity in
extracellular space can lead to low-pass frequency filtering, but this approach
was not satisfactory because high-pass filters could also be obtained, in
contradiction to experiments.  In addition, this model predicted frequency
attenuation which was not quantitative, as action potentials were still visible
at 1~mm distance, which is in contrast to what is observed experimentally.

In the present paper, we go one step further and consider an explicit structure
of extracellular space, in which we study the interaction of the electric field
with the membranes surrounding neurons.  Neurons are sourrounded by densely
packed membranes of other neurons and glial
cells~\cite{Cajal,Braitenberg98,Peters91}, a situation which we approximate here
by considering a series of passive spherical membranes around the source, all
embedded in a conducting fluid.  We show that low-pass frequency filtering can be
determined by the membranes of such passive cells and the phenomenon of electric
polarisation.

\section{General theory}
\label{sec:GeneralTheory}

In this section, we describe the theoretical framework of the model.  We start by
outlining the model (Section~[\ref{sec:ExtraCellSpace}]), where a simplified
structure of extracellular space is considered.  We also describe its main
simplifications and assumptions.  Next, we discuss the physical implementation of
this model (Section~[\ref{sec:DynamicalEqs}]), while the propagation of the
electric fields will be analyzed more formally in the next sections
(Sections~[\ref{sec:AsympBehReg1}--\ref{sec:TimeDep}]).

\subsection{Model of extracellular space}
\label{sec:ExtraCellSpace}

In Ref.\cite{Bedard04}, we considered a model where the electrical properties of
extracellular space are described by two parameters only: conductivity $\sigma$
and permittivity $\epsilon$. Both, $\sigma$ and $\epsilon$ were considered to
vary with location according to some ad hoc assumptions.  $V_{\omega}$, the
frequency $\omega$ component of the electric potential $V$ obtained by a Fourier
transformation of the potential as function of time, was shown to obey the
equation
\begin{equation} 
\label{eq:CompPot} 
\Delta V_{\omega} + (\vec{\nabla} V_{\omega}) \cdot
(\frac{\vec{\nabla} (\sigma + i \omega \epsilon)}{(\sigma + i
\omega \epsilon)}) = 0 ~ .
\end{equation}
The physical behavior of the solution is essentially determined by the expression
$1 + i \omega \frac{\epsilon}{\sigma}$. When a strong inhomogeneity occurs, one
may have $\omega \frac{\epsilon}{\sigma} >> 1$. This means that a strong phase
difference may exist between electric current and potential, i.e.  a large
impedance. In a neurophysiological context, such behavior is likely to occur at
the interface between the extracellular fluid (high conductivity) and the
membranes of cells (low conductivity).  This situation was considered previously
in the context of a simplified representation of extracellular space, in which
the inhomogeneities of conductivity were assumed to be of spherical symmetry
around the source~\cite{Bedard04}.  Moreover, the equation above assumes that the
charge density of the extracellular medium is zero when its potential is zero,
which is not entirely true since there is an excess of positive charges on the
exterior surface of neuronal membranes at rest.

In the present work, we consider an explicit structure of extracellular space to
more realistically account for these inhomogeneities of conductivity.  The
extracellular space is assumed to be composed of active cells producing current
sources (neurons), and passive cells (glia), all embedded in a conducting fluid. 
Neurons are characterized by various voltage-dependent and synaptic ion channels,
and they will be considered here as the sole source of the electric field in
extracellular space.   On the other hand, glial cells are very densely packed in
interneuronal space, sometimes surrounding the soma or the dendrites of
neurons~\cite{Peters91,Cajal}.  Glial cells normally do not have dominant
voltage-dependent channel activity, and they rather play a role in maintaining
extracellular ionic concentrations.  Like neurons, they have an excess of
negative charges inside the cell, which is responsible for a negative resting
potential (for most central neurons, this resting membrane potential is around
-60 to -80~mV).  They will be considered here as ``passive'' and representative
of all non-neuronal cell types characterized by a resting membrane potential.  We
will show that such passive cells can be polarized by the electric field produced
by neurons.  This polarization has an inertia and a characteristic relaxation
time which may have important consequences to the properties of propagation of
local field potentials.  These different cell types are separated by
extracellular fluid, which plays the role of a conducting medium, i.e. allows for
the flow of electric currents.  In the remainder of this text, we will use the
term ``passive cell'' to represent the various cell types around neurons, but
bearing in mind that they may represent other neurons as well.

Another simplification is that we will consider these passive cells as of
elementary shape (spherical or cubic).  Under such a simplification, it will be
possible to treat the propagation of field potentials analytically and design
simulations using standard numeric tools.  Our primary objective here is to
explore one essential physical principle underlying the frequency-filtering
properties of extracellular space, based on the polarization of passive membranes
surrounding neuronal sources.  We assume that such a principle will be valid
regardless of the morphological complexity and spatial arrangement of neurons and
other cell types in extracellular space.  As a consequence of these
simplifications, the present work does not attempt to provide a quantitative
description but rather an exploration of first principles that could be applied
in later work to the actual complexity of biological tissue.

\begin{figure*}[ht]
\begin{center}
\includegraphics[scale=0.7,angle=0]{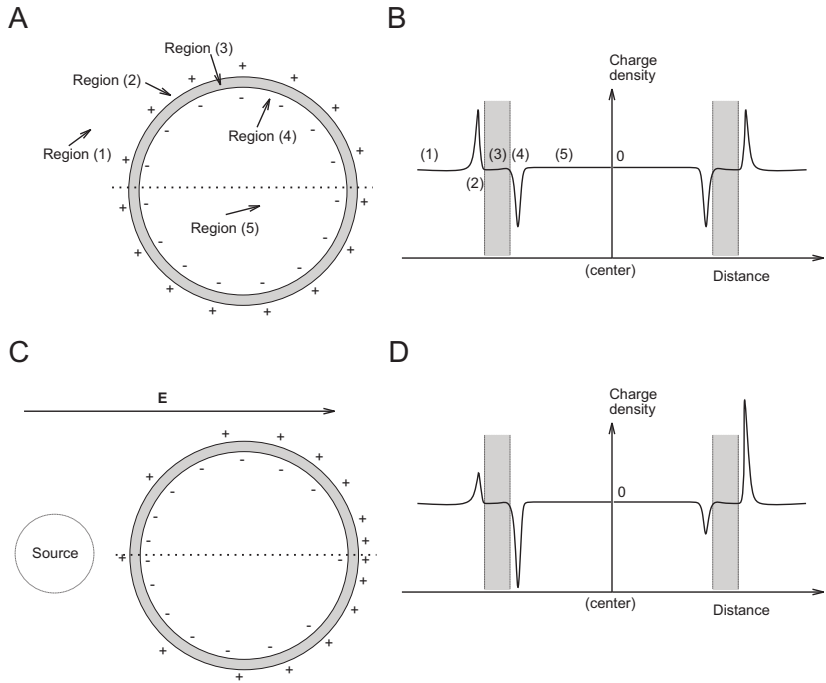}
\end{center}
\caption{Scheme of charge distribution around the membrane of a passive cell.
A. Charge distribution at rest.  The following regions are defined: the
extracellular fluid (region 1), the region immediately adjacent to the exterior
of the membrane where positive charges are concentrated (region 2), the membrane 
(region 3, in gray), the region immediately adjacent to the interior of the 
membrane where negative charges are concentrated (region 4), and the
intracellular (cytoplasmic) fluid (region 5).  B. Schematic representation of
the charge density as a function of distance (along the horizontal dotted line
in A).  C. Redistribution of charges in the presence of an electric field.  The
ions move away from or towards the source, according to their charge, resulting
in a polarization of the cell.  D. Schematic representation of the charge density 
predicted from C.}
\label{fig:charge}
\end{figure*}

The arrangement of charges in our model is schematized in
Fig.~[\ref{fig:charge}A], where we delimited 5 regions.  The membrane of the
passive cell (region 3) separates the intracellular fluid (region 5) from the
extracellular fluid (region 1), both of which are electrically neutral. The
negative charges in excess in the intracellular medium agglutinate in the region
immediately adjacent to the membrane (region 4), while the analogous region at
the exterior surface of the membrane (region 2) contains the positive ions in
excess in the extracellular space.  This arrangement results in a charge
distribution (schematized in Fig.~[\ref{fig:charge}B]) which creates a strong
electric field inside the membrane and a membrane potential.

The behavior of such a system depends on the values of conductivity and
permittivity in these different regions (they are considered constant within each
region). The extracellular fluid (region 1) has good electric conductance
properties. We have taken as conductivity $\sigma_1 = 4 ~ \mbox{Ohm}^{-1}
\mbox{meter}^{-1}$, consistent with biological data $\sigma = 3.3 - 5 ~
\mbox{Ohm}^{-1} \mbox{meter}^{-1}$, taken from measurements of specific impedance
of rabbit cerebral cortex \cite{Ranck63}. This value is comparable to the
conductivity of salt water ($\sigma_{sw} = 2.5$ Ohm$^{-1}$ meter$^{-1}$).  The
permittivity is given by $\epsilon_1$ =70 $\epsilon_{0}$, corresponding to salt
water. Here $\epsilon_{0} = 8.854 \times 10^{-12}$ Farad/meter denotes the
permittivity of the vacuum. In region 2, to the our best knowledge, there are no
experimental data on conductivity close to the membrane.  We have chosen the
values of $\sigma_2 = 0.7 \times 10^{-7} ~ \mbox{Ohm}^{-1} \mbox{meter}^{-1}$ and
$\epsilon_2 = 1.1 \times 10^{-10} ~ \mbox{Farad~meter}^{-1} \approx 12
~\epsilon_{0}$ for region 2. Such a choice is not inconsistent with biological
observations. First, electron microscopic photographs taken from the region near
the membrane reflect very little light, which hints to quite low conductivity
compared to the conductivity of region 1. We consider it as plausible that
permittivity in region 2 should be smaller than in region 1. Our choice of
$\sigma_2$ and $\epsilon_2$ corresponds to a Maxwell time $T_{M}$ yielding a
cut-off frequency $f_{c} \approx 100~Hz$, which was also the choice given in a
previous study investigating composite materials \cite{Orlowska}.

For passive cells, we neglect ion channels and pumps located in the membrane,
which is equivalent to assume the absence of any electric current across the
membrane.  Therefore, Region 3 has zero conductivity perpendicular to the
membrane surface.  The capacity of a cellular membrane has been measured and is
about $C=10^{-2}$~Farad/meter$^{2}$ \cite{Johnston99}. Approximating the membrane
by a parallel plates capacitor (with surface S and distance d, obeying
$C=\epsilon S/d$), one estimates the electric permittivity of membrane to
$\epsilon_3=10^{-10}$ Farad/meter.  Hence we used the parameters $\sigma_3 = 0$,
and $\epsilon_3 = 12 ~ \epsilon_{0}$.  

Thus, the basic idea behind the model is as follows.  As represented in
Fig.~[\ref{fig:charge}], we consider a single spherical passive cell under the
influence of an electric field.  The electric field will induce a polarisation of
the cell by reorganizing its charge distribution (Fig.~[\ref{fig:charge}C-D]). 
This polarisation will create a secondary electric field, with field lines
connecting those opposite charges. It is customary notation to call the original
electric field the {\it source field}, or the {\it primary field}, while the
field due to polarisation is called the {\it induced field}, or the {\it
secondary field}. The physical electric field is the sum (in the sense of
vectors) of both, the source and the induced field.  This induced field will be
highly dependent on frequency, for high frequencies, the ``inertia'' of charge
movement in regions 2 and 4 will limit such a polarisation, and will reduce the
effect of the induced field.  This phenomenon is the basis of the model of
frequency-dependent local field potentials presented in this article.

\subsection{Physical implementation of the model}
\label{sec:DynamicalEqs}

We introduce here the toolbox of physics used for the construction of our model,
in which we consider neurons as the principal sources of electric current. In
electrodynamics, one distinguishes between a current source and a potential
source. A current source, e.g. in an electric circuit, means that if the electric
current is used to do some work, the source maintains its level of current. 
Likewise, a potential source maintains its electric potential.  Although neurons
are generally considered as current sources, here we have chosen to consider them
as potential sources, for the following reasons.  First, these two types of
sources are equivalent for calculating extracellular potentials.  The electric
potential within a given region $D$ depends on the limit conditions at the border
of this region, and not on the type of source (current vs.\ potential source). 
Second, it is much easier to calculate the potential using limit conditions on
the potential (Dirichlet conditions) than using limit conditions with currents
(Neumann conditions).  Third, potentials are better constrained from experiments
and their range of values is better known than currents (for example the
amplitude of membrane potential variations in neurons is of the order of 10-20~mV
for subthreshold activity, and of about 100~mV during spikes).

Let us suppose that we are given a source of electric field, representing a
neuron with some open ion channels. The motion of ions through those channels
gives a current density, and also creates a distribution of charges (ions). 
According to the Maxwell equations, a given distribution of currents and charges
(plus information on the polarisation and induction properties of the medium)
determines the magnetic field, the electric field and hence the electric
potential. In a first step, we consider the effect of the electric field on a
single passive cell.

The source electric field is the origin of two physical phenomena. First, the
electric field exerts a force on charged particles (ions) and thus creates a
motion of charge carriers, i.e. an electric current. Second, the electric field
creates a displacement of charges in the borderline region of the membrane of the
passive cell (region~2). This displacement creates an induced potential due to
polarisation.  However, this polarisation is not instantaneous, due to the
``inertia'' of charge movements.  The charges on the membrane move relatively
slowly, which will be responsible for a slow time dependence of the polarisation,
before reaching equilibrium.  The characteristic time scale of such a process is
given by Maxwell's relaxation time $T_{M}$ (see Eq.~[\ref{eq:MaxwellRelaxTime}]
below), which depends on the properties of the medium, like resistivity and
permittivity.  The temporal behavior of the source is dynamic, which can
characterized by some characteristic time scale $T_{S}$. E.g., during the
creation of an action potential, ion channels open and ionic currents flow, the
electric field and potential changes. After a certain time the neuron source goes
back to its state of rest. This means that the electric current will vanish after
some delay. For example, in a typical neuron, the width of the action potential
is typically of the order of 2 msec. The polarisation and relaxation dynamics of
the passive cell depends on both time scales, $T_{S}$ and $T_{M}$. This
situation holds in the case of an ideal dielectric medium (conductivity zero, for
example pure water). In the extracellular medium, however, conductivity is
non-zero.  As a consequence, the polarisation potential and the current
distribution will mutually influence each other.  

Thus, we face the question: How do we quantitatively calculate the electric field
and potential in such a medium? And what is its time-dependence and
frequency-dependence? Looking at the temporal behavior, there are two different
regimes from the biological point of view. First, there is the transitory regime
which describes the short period of opening and shutting down of the source and
its response of the medium with some delay. Second, there is the so-called
asymptotic or permanent regime, where no current flows.  These regimes will
be considered in the next sections.

\subsection{Asymptotic behavior in region 1}
\label{sec:AsympBehReg1}

Let us start by considering the behavior in region 1 when the system has settled
into an asymptotic regime.  It means that charges move and after some time attach
to the surface of the passive cell membrane (region 2). We call this a stationary
or equilibrium regime.  In this region, the electric properties are characterized
by permittivity $\epsilon_{1}$ and conductivity $\sigma_{1}$.  We assume that
these electric properties are identical everywhere inside region 1, i.e.
permittivity $\epsilon_{1}$ and conductivity $\sigma_{1}$ are constant.  

To describe this situation, we start by recalling the fundamental equations of
electrodynamics. Gauss' law, which relates the electric field $\vec{E}$ to the 
charge density $\rho$, reads 
\begin{equation}
\label{eq:GaussLaw}
\vec{\nabla} \cdot (\epsilon_{1} \vec{E} ) = \rho ~ .
\end{equation}
Moreover, because $\sigma_{1}$ is constant, there is Ohm's law, which relates 
the electric field $\vec{E}$ to the current density $\vec{j}$,
\begin{equation}
\label{eq:OhmsLaw}
\vec{j}=\sigma_{1} \vec{E} ~ .
\end{equation}
 From Maxwell's equations one obtains the continuity equation
\begin{equation}
\label{eq:Continuity}
\vec{\nabla}\cdot \vec{j} + \frac{\partial \rho}{\partial t} = 0 ~ .
\end{equation}
Using Eqs.~[\ref{eq:GaussLaw}-\ref{eq:Continuity}] and recalling that 
$\epsilon_{1}$ and $\sigma_{1}$ are both 
constant in region 1 implies the following differential equation for the 
charge density,
\begin{equation}
\label{eq:Density}
\frac{\partial \rho}{\partial t} = - 
\frac{\sigma_{1}}{\epsilon_{1}} \rho ~ .
\end{equation}
A particular solution to this equation requires to specify boundary conditions
and initial conditions.
Here the boundary conditions are such that the electric field (the component
of the field perpendicular to the surface) on the surface of the source and 
on the surface of the passive cell is given (Dirichlet boundary conditions). 
The general solution of Eq.~[\ref{eq:Density}] is
\begin{eqnarray}
\label{eq:RhoExpLaw}
&\rho(\vec{x},t)& = \rho(\vec{x},0) \exp[ - \frac{\sigma_{1}}{\epsilon_{1}} t ]
= \rho(\vec{x},0) \exp[ - t/T_{M_1} ] ~ ,
\nonumber \\
&\rho(\vec{x},t)& \longrightarrow_{t \to \infty} 0 ~ ,
\end{eqnarray}
i.e., with increasing time the charge distribution goes exponentially to zero. 
The time scale, which characterizes the exponential law, is Maxwell's time of 
relaxation, $T_{M_1}$. In general it is defined by
\begin{equation}
\label{eq:MaxwellRelaxTime}
T_{M} = \frac{\epsilon}{\sigma} ~ .
\end{equation}
In particular, in region 1, one has $T_{M_1}=\epsilon_{1}/\sigma_{1}$.  In the
limit of large time, the continuity equation implies $\vec{\nabla} \cdot \vec{j}
= 0$, i.e. there are no sources or sinks.  Also in this limit the electric
potential satisfies Laplace's equation, $\Delta V = 0$, which follows from Gauss'
law. Maxwell's relaxation time $T_{M_1}$ in region 1 is very short, of the
order of $10^{-10}$~s, thus the charge density tends very quickly to zero, and so
does the extracellular potential.

\subsection{Behavior in region 2}
\label{sec:AsympBehReg2}

Region 2 is the near neighborhood of the membrane of a passive cell.  The
neuronal source creates the source field (or primary field) $\vec{E}_{source}$. 
As pointed out above (Fig.~[\ref{fig:charge}C]), due to the presence of a source
and the presence of free charges near the passive cell membrane, there will be a
polarisation of the free charges, described by an electric induced field (or
secondary field) $\vec{E}_{ind}$ also denoted by $\vec{E}_{free}$.  Let us assume
that the source field is "switched on" at time $t=0$. Such time-dependence is
described by a Heaviside step function $H(t)$ (see Eq.~[\ref{eq:AppC-2}] in the
Appendix).
\begin{equation}
\vec{E}_{source}(\vec{x},t) = \vec{E}_{0}(\vec{x}) ~ H(t) ~ .
\end{equation}
The electric field present in region 2 results from the source field 
$\vec{E}_{source}$, the field due to free charges 
$\vec{E}_{free}$ (which create the induced field) and the field due to fixed 
localized charges (dipoles) of the membrane $\vec{E}_{membr}$. 
Under the hypothesis that the passive cell membrane is a rigid structure 
with dipoles in fixed locations and under the assumption that 
ion channels in that membrane remain closed, 
we conclude that the electric field $\vec{E}_{membr}$ does not vary in time.
Gauss' law and the continuity equation now read
\begin{eqnarray}
\vec{\nabla} \cdot \vec{E}_{free} = \frac{\rho_{free}}{\epsilon_{2}} ~ ,
\nonumber \\
\vec{\nabla} \cdot \vec{j} = 
- \frac{\partial \rho_{free}}{\partial t} ~ .
\end{eqnarray}
Ohm's law reads
\begin{equation}
\vec{j} = \sigma_{2}(\vec{E}_{source} + \vec{E}_{free} + \vec{E}_{membr}) ~ ,
\end{equation}
which implies
\begin{equation}
\label{eq:EqMotFreeCharge}
\frac{1}{\sigma_{2}} \frac{\partial \rho_{free}}{\partial t} 
= - \vec{\nabla} \cdot (\vec{E}_{source} + \vec{E}_{free} + \vec{E}_{membr}) 
= - \frac{\rho_{free}}{\epsilon_{2}} + f(\vec{x}) ~~~ \mbox{for} ~ t > 0 ~ . 
\end{equation}
While $\rho_{free}$ depends on position and time, the function $f(\vec{x})$ 
denotes a time-independent term ($\vec{E}_{membr}$ is time-independent and 
$\vec{E}_{source}$ is time-independent for $t > 0$). The solution of 
Eq.~[\ref{eq:EqMotFreeCharge}] becomes
\begin{equation}
\label{eq:SolFreeCharge}
\rho_{free}(\vec{x},t) = a(\vec{x}) + b(\vec{x}) \exp[ - t/T_{M_2} ] 
~~~ \mbox{for} ~ t > 0 ~ .
\end{equation}
Here, $T_{M_2}=\epsilon_{2}/\sigma_{2}$ denotes the Maxwell time in region 2.
The function $a(\vec{x})$ represents the free charge density at equilibrium, that
is a long time ($t=\infty$) after the source has been switched on and the free
charges have settled in region 1 in such a way that in region 1 no net electric
field is left and hence no flow of electric current occurs.  The function
$b(\vec{x})$ denotes the difference of the free charge density at the moment when
switching on the source and the free charge density at equilibrium. One should
note that $\rho_{free}(\vec{x},t)$ is a continuous function in $\vec{x}$ at
$t=0$.  Poisson's equation implies that a linear relation holds between the free
charge density and the induced potential. This and Eq.~[\ref{eq:SolFreeCharge}]
yields 
\begin{equation}
\label{eq:SolIndPot}
V_{ind}(\vec{x},t) = c(\vec{x}) + d(\vec{x}) \exp[ - t/T_{M_2} ] 
~~~ \mbox{for} ~ t > 0 ~ .
\end{equation}
In order to understand the meaning of the functions 
$c(\vec{x})$ and $d(\vec{x})$, let us consider as example 
the following source potential
\begin{equation}
\label{eq:SourcePot}
V_{source}(\vec{x},t) = \beta H(t) + \alpha [1-H(t)] ~ ,
\end{equation}
where $\alpha$ and $\beta$ are taken as constant in space and time and $H(t)$ is 
the Heaviside function defined in the Appendix. This source function makes a 
jump immediately after $t=0$. Then the induced potential becomes
\begin{equation}
\label{eq:IndPot}
V_{ind}(\vec{x},t) = V_{ind}^{equil}(\vec{x}) + 
[V_{ind}(\vec{x},t=+\delta) - V_{ind}^{equil}(\vec{x})] 
\exp[ - t/T_{M_2} ] 
~~~ \mbox{for} ~ t > 0 ~ ,
\end{equation}
where $V_{ind}^{equil}(\vec{x})$ denotes the induced potential at equilibrium 
and $V_{ind}(\vec{x},t=+\delta)$ denotes the induced potential immediately after 
the source has been switched on. The resulting total potential is given by
\begin{equation}
\label{eq:ResultPot}
V(\vec{x},t) = V_{source}(\vec{x},t=+\delta) + V_{ind}(\vec{x},t)  
~~~ \mbox{for} ~ t > 0 ~ ,
\end{equation}
where $V_{source}(\vec{x},t=+\delta)$ denotes the source potential, 
Eq.~[\ref{eq:SourcePot}], immediately after the source has been switched on.

Now let us consider more specifically a source with time-dependence given by 
the following function,
\begin{equation}
V_{source}(\vec{x},t) = \left\{
\begin{array}{ll}
\beta H(t) & \mbox{case} (a) ~ , \\
\alpha [1-H(t)] & \mbox{case} (b) ~ . 
\end{array}
\right.
\end{equation}
In case (a), the source potential is constant ($=\beta$) in space and jumps in
time from zero to one immediately after $t=0$, meaning the source is switched on.
In case (b), the source potential is constant ($=\alpha$) in space and jumps in
time from one to zero immediately after $t=0$, meaning the source is switched
off. By repetition of alternate switch-on's and switch-off's, one can introduce a
temporal pattern with a certain frequency, as shown in
Fig.~[\ref{fig:squarewave}A]. Here we are interested in the temporal behavior of
the induced potential under such circumstances. In case (a)
Eq.~[{\ref{eq:IndPot}] implies for the induced potential the following relation
\begin{equation}
\label{eq:IndPotCase:a}
V_{ind}(\vec{x},t) = V_{ind}^{equil}(\vec{x}) 
[1 - \exp[ - t/T_{M_2} ]] 
~~~ \mbox{for} ~ t > 0 ~ .
\end{equation}
By differentiation, one finds that the induced potential obeys the following 
differential equation,
\begin{equation}
\label{eq:DiffEqCase:a}
\frac{\partial V_{ind}(\vec{x},t)}{\partial t} + 
\frac{1}{T_{M_2}} V_{ind}(\vec{x},t) = 
\frac{1}{T_{M_2}} V_{ind}^{equil}(\vec{x}) ~ .
\end{equation}
Similarly, in case (b) one obtains for the induced potential
\begin{equation}
\label{eq:IndPotCase:b}
V_{ind}(\vec{x},t) = V_{ind}(\vec{x},t=o) \exp[ - t/T_{M_2} ] 
~~~ \mbox{for} ~ t > 0 ~ ,
\end{equation}
which obeys the following differential equation,
\begin{equation}
\label{eq:DiffEqCase:b}
\frac{\partial V_{ind}(\vec{x},t)}{\partial t} + 
\frac{1}{T_{M_2}} V_{ind}(\vec{x},t) = 0 ~ .
\end{equation}
Eqs.~[\ref{eq:DiffEqCase:a},\ref{eq:DiffEqCase:b}] can be expressed as
\begin{eqnarray}
\label{eq:DiffEqBothCases}
&& \frac{\partial V_{ind}(\vec{x},t)}{\partial t} + 
\frac{1}{T_{M_2}} V_{ind}(\vec{x},t) = 
\frac{1}{T_{M_2}} f(\vec{x}) H(t) ~ ,
\nonumber \\
&& f(\vec{x}) = \left\{ 
\begin{array}{cc}
V_{ind}^{equil}(\vec{x}) & \mbox{case} (a) \\
0 & \mbox{case} (b)
\end{array}
\right. ~ .
\end{eqnarray}
In order to solve this differential equation, one has to know the function 
$f(\vec{x})$ given by $V^{equil}_{ind}(\vec{x})$ in case (a).
How to obtain this function? We consider the case of an ideal dielectric 
medium in region 1. The function $V^{equil}_{ind}(\vec{x})$ is related to the 
full potential and the source potential via
\begin{equation}
V^{equil}_{ind}(\vec{x}) = V^{equil}_{tot} - V^{equil}_{source}(\vec{x})
\end{equation}
The resulting total potential at equilibrium $V^{equil}_{tot}$ on the surface 
of the membrane will be constant in time and in space (on the membrane surface), 
i.e. independent of position $\vec{x}$, because otherwise there would be a flow 
of charges. The value of $V^{equil}_{tot}$ can be found by computing 
\begin{equation}
\label{eq:GaussTheorem}
\int_{S} d\vec{S} \cdot \epsilon_1 \vec{E}^{equil}_{tot} = Q^{equil}_{tot}
\end{equation}
The integral is done over any surface $S$ in the extracellular fluid, chosen such
that it only includes the passive cell. $Q^{equil}_{tot}$ is the total charge in
the interior of such surface $S$. The physical value of $V^{equil}_{tot}$ must be
such that the corresponding electric field $\vec{E}^{equil}_{tot}$ yields
$Q^{equil}_{tot}=0$ via Eq.~[\ref{eq:GaussTheorem}], because the total charge of
the passive cell before switching on the source was neutral.  

Note that the above reasoning is valid for an ideal dielectric medium, which is
not the case for extracellular media.  However, the small amplitude of the
currents involved ($\sim$ 100~pA), the value of conductivity of extracellular
fluid ($\sim$~3.3~S/m), the small dimension of most passive cells
($\sim$~10~$\mu$m diameter; $\sim$~2~nm of membrane thickness), and the high
resistivity of membranes, imply a weak voltage drop on cell surfaces due to the
current.  Thus, the electrostatic induction is very close to an ideal dielectric.

\subsection{Source given by periodic step function:
an equivalent RC electric circuit}
\label{sec:AnalogRC-Circuit}

We considered above initial conditions where the source potential has been
switched on at some time.  This can be directly generalized such that the
temporal behavior of the source potential is given by a periodic step function,
with a period $T_{S}$, as shown in Fig.~[\ref{fig:squarewave}A]. The
corresponding induced potential, presented in Fig.~[\ref{fig:squarewave}B], shows
a piecewise exponential increase followed by an exponential decrease.
\begin{figure*}[ht]
\begin{center}
\includegraphics[scale=0.57,angle=0]{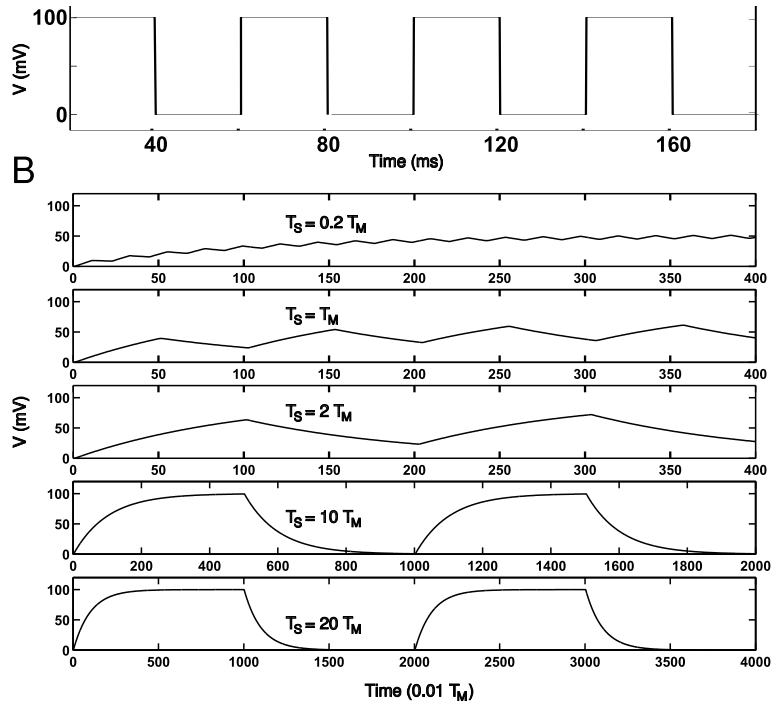}
\end{center}
\caption{
Time-dependence of induced electric potential 
in response to an external source given by a periodic function.
A. Source potential given by a periodic step function $H(sin(\omega t))$, with 
period $T_{S} = 2 \pi/\omega = 80 ms$.
B. Induced electric potential for various values of 
$T_{S}/T_{M}$.
In the case $T_{S}/T_{M} << 1$ (top), the induced potential fluctuates closely 
around the mean value of the source potential. In the case 
$T_{S}/T_{M} >>1 $ (bottom) the induced potential relaxes and 
fluctuates between $V_{min}=0~mV$ and $V_{max}=100~mV$ of the source potential.
As a result, the amplitude of oscillation becomes more attenuated with 
increased frequency $f=1/T_{S}$.
}
\label{fig:squarewave}
\end{figure*}
The figure shows the response to a source of periodic step function, for
different values of $T_{S}$ measured in units of $T_{M}$. One observes
the following behavior. At the top of the figure the period $T_{S}$ is
smallest, i.e. which corresponds to a rapid oscillation about its time average.
At the bottom of the figure the period $T_{S}$ is largest, corresponding to
a low-frequency oscillation. It is important to note that the amplitude of
oscillation (i.e. the difference between its maximal and minimal value) is much
smaller for small $T_{S}$ than for large $T_{S}$. This yields an
attenuation effect of high frequencies in the induced potential.

Such behavior of the induced potential is well known from and mathematically
equivalent to that of an RC electric circuit, with a resistor $R$ and a 
capacitor $C$ (see Fig.~[\ref{fig:RC}]).  
\begin{figure*}[ht]
\begin{center}
\includegraphics[scale=0.40,angle=0]{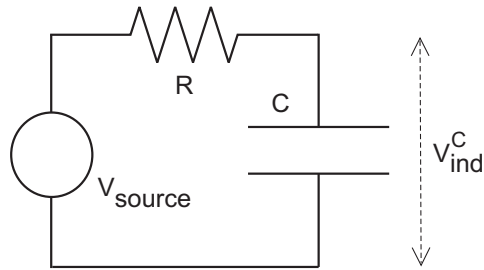}
\end{center}
\caption{Model of electric circuit with capacitor $C$, resistor $R$. 
$V_{source}$ denotes the potential of the source and $V_{ind}^{C}$ denotes the 
induced potential at the capacity $C$. }
\label{fig:RC}
\end{figure*}
The equation of motion in such an $RC$ circuit relating the induced potential 
$V_{ind}^{C}$ to the capacity $C$ to the source potential $V_{source}$ is given by
\begin{equation}
\label{eq:EqMotionElecCircuit}
\frac{\partial V_{ind}^{C}(t)}{\partial t} + \frac{1}{RC} V_{ind}^{C}(t) = \frac{1}{RC} V_{source}(t) ~ .
\end{equation}
This equation is mathematically equivalent to Eq.~[\ref{eq:DiffEqBothCases}], if 
we identify
\begin{eqnarray}
\label{eq:Correspondance}
&& V_{ind}(\vec{x},t) \longleftrightarrow V_{ind}^{C}(t) ~ ,
\nonumber \\
&& V_{ind}^{equil}(\vec{x}) H(t) 
\longleftrightarrow V_{source}(t) ~ ,
\nonumber \\
&& \frac{1}{\sigma} \longleftrightarrow R ~ ,
\nonumber \\
&& \epsilon \longleftrightarrow C ~ ,
\nonumber \\
&& T_{M} \longleftrightarrow RC ~ .
\end{eqnarray}
Thus, for each point $\vec{x}$ near the membrane of a passive cell (region 2 in
Fig.~[\ref{fig:charge}A]), we can set up an equivalent electric circuit, which is 
equivalent in the sense that it gives a differential equation with the same 
mathematical solution as for the induced potential of the original system of a 
source and a passive cell.

\subsection{Induced potential for time dependent source}
\label{sec:TimeDep}

The analogue of an RC electric circuit can be used to determine the
characteristic properties of the source/passive-cell system, like the cut-off
frequency.  For that purpose let us consider the source potential to be given by
a periodic function of sinusoidal type switching on at time $t=0$:
\begin{equation}
V_{source}(\vec{x},t) = V_{0}(\vec{x}) \ \exp(i \omega t) \ H(t)  ~ .
\end{equation}
The asymptotic (large $t$) solution of the induced potential is given by
\begin{equation}
V_{ind}(\vec{x},t) \sim_{t \to \infty} 
\frac{\sigma}{\sigma + i \omega \epsilon} V_{0}(\vec{x}) \exp(i \omega t) ~ .
\end{equation}

Fig.~[\ref{fig:squarewave}B] shows an example of the induced potential, 
where the source has a time dependence given by an oscillating step function 
(Fig.~[\ref{fig:squarewave}A]).  
We use the notion of the transfer function, which denotes the ratio of induced 
potential corresponding to frequency $f$ of the source over the induced 
potential at frequency $f=0$ of the source,
\begin{equation}
\label{eq:DefTransferFct}
F_{TM}(\vec{x},\omega) = \lim_{t \to \infty} 
\frac{ V_{ind}^{\omega}(\vec{x},t) \exp(-i \omega t) }
{ V_{ind}^{\omega=0}(\vec{x},t) } ~ .
\end{equation}
Here, $V_{ind}^{\omega}(\vec{x},t)$ denotes the solution of the differential 
equation [\ref{eq:AppC-1}] with a source term having an asymptotic 
time-dependence given by $\exp(i \omega t)$. 
There is a variety of methods to analyze the time-dependence of induced 
potentials for a given time-dependent source.   
The method of Fourier analysis, where the time dependence of a 
function is decomposed in terms of function $\exp(i\omega t)$, 
has been proven very useful in the time-analysis 
of experimental data (e.g. the power spectrum). 
An alternative method is the Heaviside method, where the 
time dependence of a function is decomposed in terms of 
oscillating step functions $H(\sin(\omega t))$. 
In the Appendix we discuss the asymptotic behavior 
of the induced potential, where the source term is given in 
terms of Heaviside step functions and also in terms of 
Fourier components. Then the transfer function for a source 
given by a Fourier component $\exp(i \omega t)$ is given by
\begin{equation}
\label{eq:TransFctOmega}
F_{TM}(\omega) = \frac{1}{1 + i \omega \epsilon /\sigma} ~ .
\end{equation}
$F_{TM}$ is a complex function, with its modulus given by
\begin{equation}
\label{eq:AbsTransFctOmega}
| F_{TM}(\omega) | = \frac{1}{\sqrt{1 + (\omega \epsilon/\sigma)^{2} } } ~ .
\end{equation}
For frequency zero one has $| F_{TM} |=1$.
The cut-off frequency $f_{c}$ is defined such that $| F_{TM} |$ 
falls off to the value $| F_{TM} | = 1/\sqrt{2}$. Thus we find 
\begin{equation}
\label{eq:CutOffFrequency}
f_{c} = \frac{\omega_{c}}{2 \pi} = \frac{\sigma}{2 \pi \epsilon} = 
(2 \pi T_{M})^{-1} ~ .
\end{equation}
The modulus of the transfer function vs. frequency is shown in
Fig.~[\ref{fig:filter}], for the case of parameters $\sigma$ and $\epsilon$ of
region 2, as given in sect.(\ref{sec:ExtraCellSpace}).  This corresponds to a 
cut-off frequency $f_{c}$ = 100~Hz. Such behavior represents a frequency filter.
It shows that we can compute the induced potential as a function of frequency
based on the analogy with an equivalent electric circuit with a resistor $R$ and
capacity $C$ arranged in serial order.  
\begin{figure*}[ht]
\begin{center}
\includegraphics[scale=0.50,angle=0]{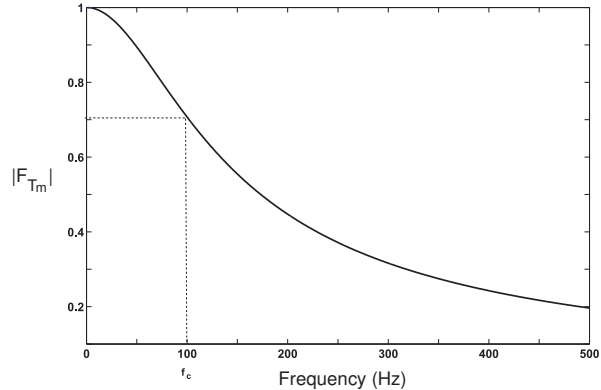}
\end{center}
\caption{Modulus of transfer function vs. 
frequency, using biologically realistic parameters $\epsilon$ and $\sigma$.}
\label{fig:filter}
\end{figure*}
A series of detailed numerical simulations, modeling frequency filtering of
extra-cellular neural tissue in the neighborhood of passive cells in terms of the
electric circuit model, Eq.~[\ref{eq:EqMotionElecCircuit}], using physiologically
realistic parameters of $\epsilon$ and $\sigma$ are shown in
Fig.~[\ref{fig:filtered-signals}].  They present the signal coming from a
periodic step function with a period $T_{S}=10~ms$, i.e. a frequency of
100~Hz. Going from Fig.~[\ref{fig:filtered-signals}A] to
Fig.~[\ref{fig:filtered-signals}C] $T_{S}$ is kept fixed but $T_{M}$
increases, i.e. $T_{S}/T_{M}$ decreases.  
Fig.~[\ref{fig:filtered-signals}A]
corresponds to the case where the relaxation time $T_{M}$ is 10 times smaller
than the period of the signal $T_{S}$.  In Fig.~[\ref{fig:filtered-signals}B]
$T_{M}$ and $T_{S}$ are equal, and in Fig.~[\ref{fig:filtered-signals}C]
$T_{M}$ is 10 times larger than $T_{S}$.  We observe that the shape of the 
signal after filtering is more or less intact for $T_{S}/T_{M}= 10$ and 
becomes gradually more deformed when going over to $T_{S}/T_{M}= 0.1$. As the 
Fourier transformation after filtering shows, the higher frequency components 
get gradually more suppressed. This represents a low frequency band pass.  
\begin{figure*}[ht]
\begin{center}
\includegraphics[scale=0.50,angle=0]{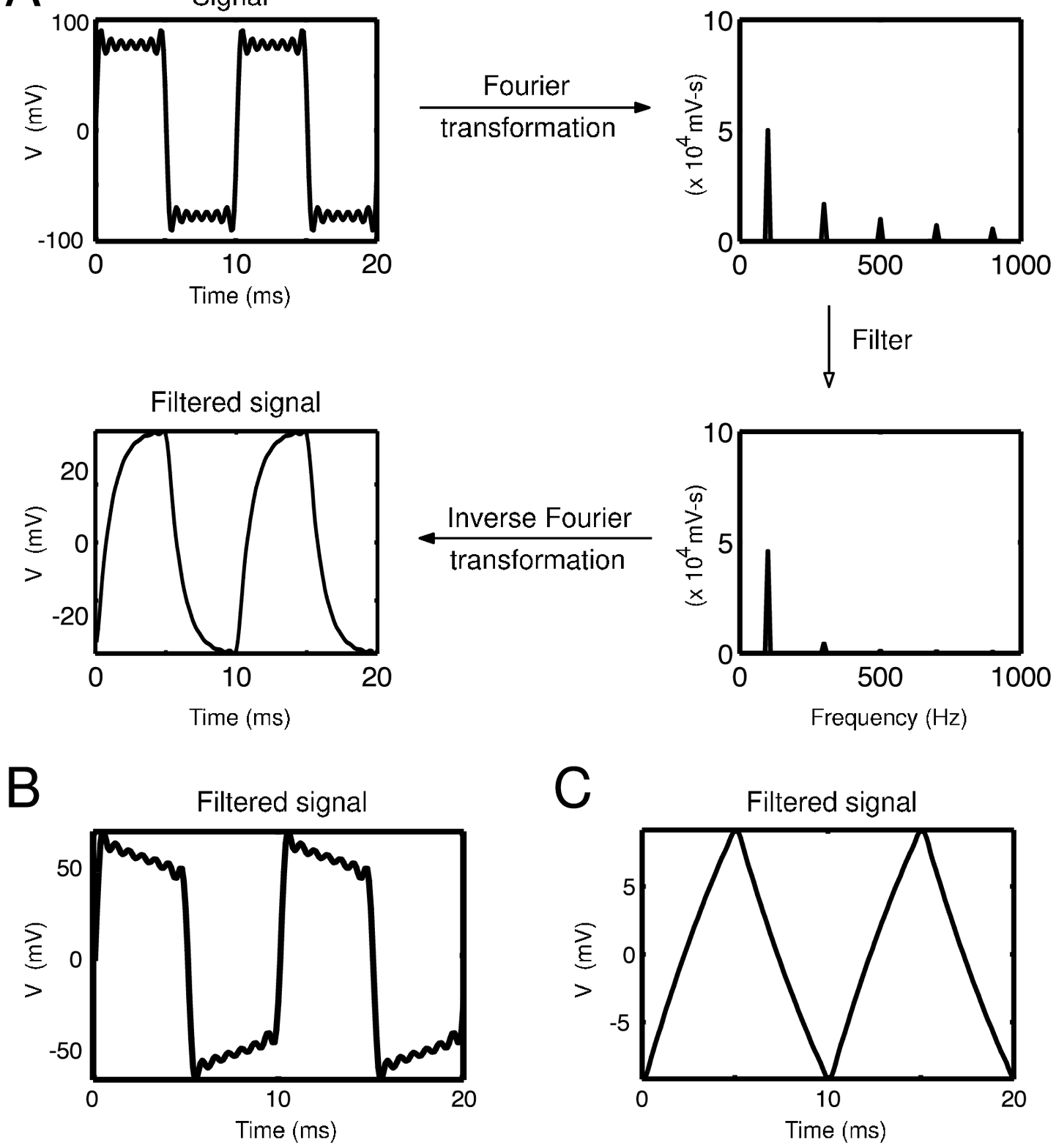}
\end{center}
\caption{Effect of filtering of a periodic electrical potential representing a neural source. 
A. The original signal (top row, left panel). The frequency spectrum of such signal is obtained by Fourier transformation (top row, right panel). Then the signal is sent through a filter (middle row, right panel). After an inverse Fourier transformation the filtered signal as function of time is obtained (middle row, left panel). 
The period of the original 
signal is 10 ms, and the relaxation time of the filter is 10 ms.
B. Same signal as in A, filtered using a relaxation time of 1 ms.
C. Same signal as in A, filtered using a relaxation time of 100 ms.}
\label{fig:filtered-signals}
\end{figure*}

To close this section, we would like to remark that the treatment above is
general and not restricted to step functions.  The Heaviside functions were used
as a tool to distinguish transient and asymptotic behaviors, as well as to
calculate the transfer function.  For an arbitrarily complex time-dependent
source, the convolution integral with this transfer function gives the filtered
signal, so this approach can be applied to any physical signal.

\section{Modeling frequency-dependence of multiple neurons and passive cells}
\label{sec:ModelMultiNeurAstro}

In Section~[\ref{sec:GeneralTheory}], we laid out the theoretical framework of
the model and showed that this model -- with respect to time and
frequency-dependence but neglecting spatial dependence -- is equivalent to a
RC-circuit model. As an example, we have presented numerical simulations treating
the case of a single source and a single passive cell. In this section, we would
like to to generalize this approach to take into account space- and
time-dependence in LFPs.  In this and the following section, we present numerical
simulations considering more complex three-dimensional arrangements of neuronal
sources and passive cells.

\subsection{Computation of field potentials for multiple neurons and passive cells}
\label{sec:ResultMultiNeurAstro}

In a first set of numerical simulations, we considered three-dimensional
arrangements of sources and passive cells.  To solve the equations, we used a
discretization method similar to finite elements methods (see below).  The neuron
source was represented by a cubic shape of length $L_{neuron}$, and likewise, we
consider the passive cells were represented by cubes of length $L_{pass}$. 
Neuron source and passive cells were located at some distance $d$ from each
other.  Although being far from realistic, this geometry facilitates numerical
simulations.  Since the aim of this work is not to make quantitative predictions,
such an arrangement will help explaining qualitatively the frequency attenuation
properties of field potentials in three-dimensional arrangements of cells, as we
will show below.

Let us first outline the method of numerical computation of the potential. From a
mathematical point of view, the solution of a partial differential equation
requires to specify suitable boundary conditions. In the context of Maxwell's
equations, Dirichlet's boundary conditions (potential given on a closed surface)
or von Neumann's boundary conditions (electric field given on a closed surface)
are known to guarantee a unique solution. Here, we suggest to use boundary
condition of Dirichlet's type, that is we assume the potential is given on the
surface of all cells. The problem is, however, that the value of the physical
potential on the surface of those cells is not known a priori, and, moreover, the
functional values of the potential can not be chosen freely. The physical
solution is determined by the principle that all free charges are located such
that the electromagnetic field energy becomes minimal (Thomson's theorem). This
imposes constraints on the boundary conditions.  

How can we figure out those a priori unknown boundary conditions?  We use the
physical principle that those physical boundary conditions are equivalent to the
condition that Gauss' integral summing the flux around a charge source becomes
invariant when the integration surface is changed. A computational strategy to
find those boundary conditions is to use a variational principle, starting from
some initial guess of boundary conditions and then to adjust iteratively the
boundary conditions until eventually the total electric energy attains its
minimum.  

We used a variational principle via the following iterative scheme.  In a first
step, we compute the potential due to the source in absence of any passive cell
in the asymptotic regime (a long time after the source has been switched on), for
the case where the source has no frequency dependence ($\omega=0$). We assume
that the potential obeys the boundary condition on the surface of the neuron
source
\begin{equation}
\label{eq:BCSource}
V(\vec{x}) = V^{0} = \mbox{const} ~~~ \mbox{for} ~ 
\vec{x} ~ \mbox{on~surface~of~source} ~ .
\end{equation}
The potential satisfies Laplace's equation
\begin{equation}
\label{eq:Laplace}
\Delta V(\vec{x}) = 0 ~ .
\end{equation}
The numerical solution has been carried out using the standard 
method of relaxation.
  
In a second step, we place a passive cell at some distance $d$ from the neuron
source. We want to calculate the induced potential due to the presence of the
passive cell. We proceed in the following way. Let us consider for a moment that
the passive cell is small compared to the size of the neuron, $L_{pass} <<
L_{neuron}$.  Then the induced potential on the surface of the passive cell would
be almost identical to $V^{center}_{pass}$, the potential created by the source
alone, evaluated at the center of the passive cell. Then imposing the boundary
condition that the potential on the surface of the passive cell takes the value
$V^{center}_{pass}$ would be very close to the exact solution. Now we consider
the passive cell no longer as tiny. Then imposing the boundary condition boundary
on the surface of the passive cell
\begin{equation}
\label{eq:BCAstro}
V(\vec{x}) = V^{center}_{pass} = \mbox{const} ~~~ 
\vec{x}~ \mbox{on~surface~of~passive cell} 
\end{equation}
becomes an approximation.

In a third step, taking into account that the potential obeys the boundary
conditions on the neuron source and on the passive cell,
Eqs.~[\ref{eq:BCSource},\ref{eq:BCAstro}], we solve again Laplace's equation,
Eq.~[\ref{eq:Laplace}].  Now we test if the obtained solution $V(\vec{x},t)$ is
the correct solution.  Using Gauss' theorem, one has
\begin{equation}
\label{eq:SurfaceInt}
\int_{S_{neuron}} d\vec{s} \cdot \vec{E} = \int_{S_{neuron+pass}} d\vec{s} \cdot \vec{E} ~ .
\end{equation}
Here the integral denotes an integral over a closed surface, first englobing the
neuron source and second englobing the neuron source plus the passive cell.
Because the passive cell has a total charge $Q_{pass}=0$,
Eq.~[\ref{eq:SurfaceInt}] should be satisfied by the physical solution
$V_{phys}(\vec{x},t)$. As long as $V(\vec{x},t)$ differs from
$V_{phys}(\vec{x},t)$, one has to adjust the potential on the surface of the
passive cell, solve again Laplace's equation and verify the charge balance
Eq.~[\ref{eq:SurfaceInt}]. This a an iterative process, which turned out to
converge quite fast towards the physical solution.  

This method can be applied as well to treat multiple neuron sources and multiple
passive cells. We would like to point out that the extracellular neural tissue is
composed of much more passive cells than neurons. On average the number of glial
cells per unit volume is higher than the number of neurons per unit volume by
about a factor of 10. The physical consequence of this property is that the
induced potential due to the presence of passive cells becomes more important
than the source potential.  

Finally, to investigate the effect of time-dependent sources and multiple passive
cells, we take advantage of the following equivalence: the induced potential on a
cell with a relaxation time $T_{M}$ subject to a time dependent source is
equivalent to the induced potential in a cell with $T_{M}=0$ subject to a
``filtered'' source given by the convolution of the source with the transfer
function ($F_{TM}$ in Eq.~[\ref{eq:DefTransferFct}]).  This corresponds to the
following calculation steps, for each passive cell: (1) evaluate the source
$S(\vec{x}, \omega)$ provided by the neighboring system of sources and other
passive cells; (2) evaluate the ``filtered'' source $S^*(\vec{x}, \omega) =
S(\vec{x}, \omega) F_{TM}(\omega)$ (from Eq.~[\ref{eq:TransFctOmega}]); (3)
calculate the induced potential on the passive cell using a similar procedure as
outlined above for a time-independent source ($\omega=0$).  The whole procedure
is repeated iteratively until all sources and induced potentials converge towards
the physical solution.  Numerical results using this iterative procedure are
presented in the next section. 

It is important to note that in the range of frequencies considered here, the
wavelengths are too large for significant wave propagation phenomena. 
Accordingly, we only consider sources where space and time dependences factorize.
Consequently, the transfer function does not have any space dependence, as shown
in the calculation in the Appendix, Eqs.~[\ref{eq:AppC-1}-\ref{eq:AppC-15}].  In
addition, the application of the transfer function holds for spherical sources as
well as for cubic sources.

\subsection{Numerical results}
\label{sec:NumResults}

We now present results of numerical solutions of Laplace's equation for the
scenario in the presence of a source and several passive cells.  We have taken as
parameters of conductivity and permittivity those corresponding to the
intracellular space at the membrane of the passive cell, given in sect. 
[\ref{sec:ExtraCellSpace}] (conductivity is non-zero). It corresponds to a
cut-off frequency of $f_{c}=100~Hz$. The corresponding Maxwell time and transfer
function are obtained by Eqs.~[\ref{eq:CutOffFrequency},\ref{eq:TransFctOmega}].  

The goal is to show that the induced potential gives an important contribution to
the total, i.e. physical potential. The results are presented in
Figs.~[\ref{fig:obst1}--\ref{fig:obst4}]. For the case of a single source and a
single passive cell the source potential, the induced potential and the total
potential are shown in Fig.~[\ref{fig:obst1}A-C], respectively.  Comparing the
source potential (B) with the induced potential (C), one observes that the latter
gives a substantial contribution. In Fig.~[\ref{fig:obst2}A-D] we show the induced
potential multiplied with the norm of the transfer function. We observe that the
frequency-dependent induced potential decreases when the frequency of the source
increases from 0 to 400~Hz.  Figs.~[\ref{fig:obst3}--\ref{fig:obst4}] show the
corresponding results in the case of a single source and four passive cells. We
observe qualitatively the same results as in Fig.~[\ref{fig:obst1}]. These results
demonstrate clearly the phenomenon of attenuation of high frequencies in the
induced potential. It should be mentioned that the computations were carried out
in 3-D, but are depicted as projection in 2-D.  Qualitatively the same results
are obtained in the case of two sources and a single passive cell (not shown). 
Again we found for the time-independent source that the induced potential is
non-negligible compared to the source potential and the frequency-dependent
induced potential rapidly decreases when the frequency goes from 0 to 400~Hz.  
\begin{figure*}[ht]
\begin{center}
\includegraphics[scale=0.75,angle=0]{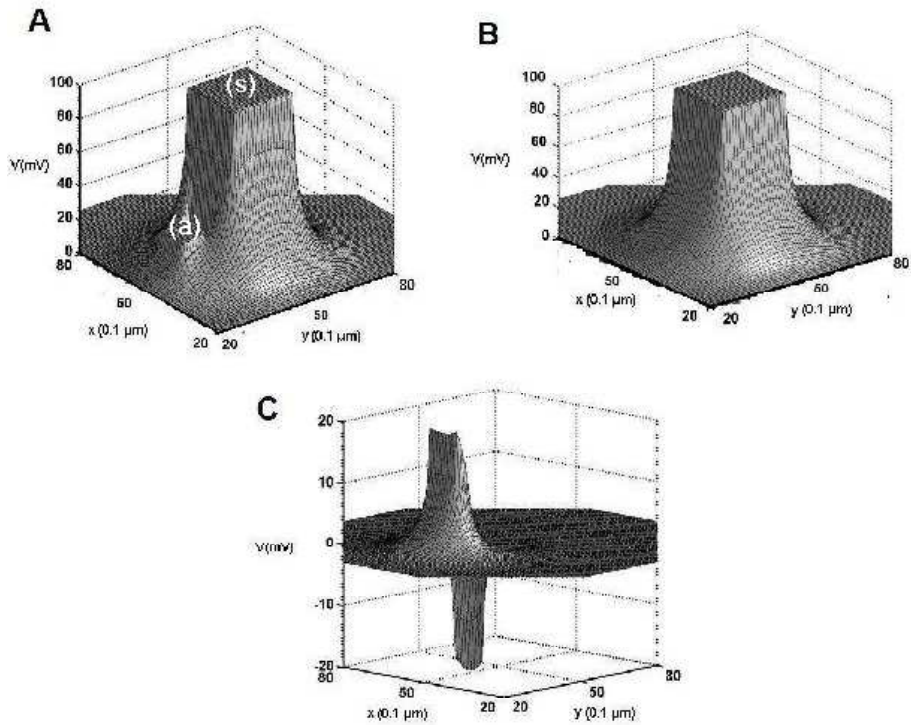}
\end{center}
\caption{Extracellular potential generated by a time-independent source situated
nearby a passive cell. Both the source (s) and the passive cell (a) have a
cubic shape; the passive cell was twice smaller than the source. A. Total
extracellular electric potential ($V_{source}+V_{ind}$) in a horizontal plane cut
at the base of both cells. B. Electric potential resulting from the source only
($V_{source}$).  C. Induced electric potential ($V_{ind}$). All potentials were
obtained by solving Laplace's equation with a static source ($f=0$), and are
represented in units of percent of the source potential. The distance is
represented in units of 0.1~$\mu m$.}
\label{fig:obst1}
\end{figure*}
\begin{figure*}[ht]
\begin{center}
\includegraphics[scale=0.85,angle=0]{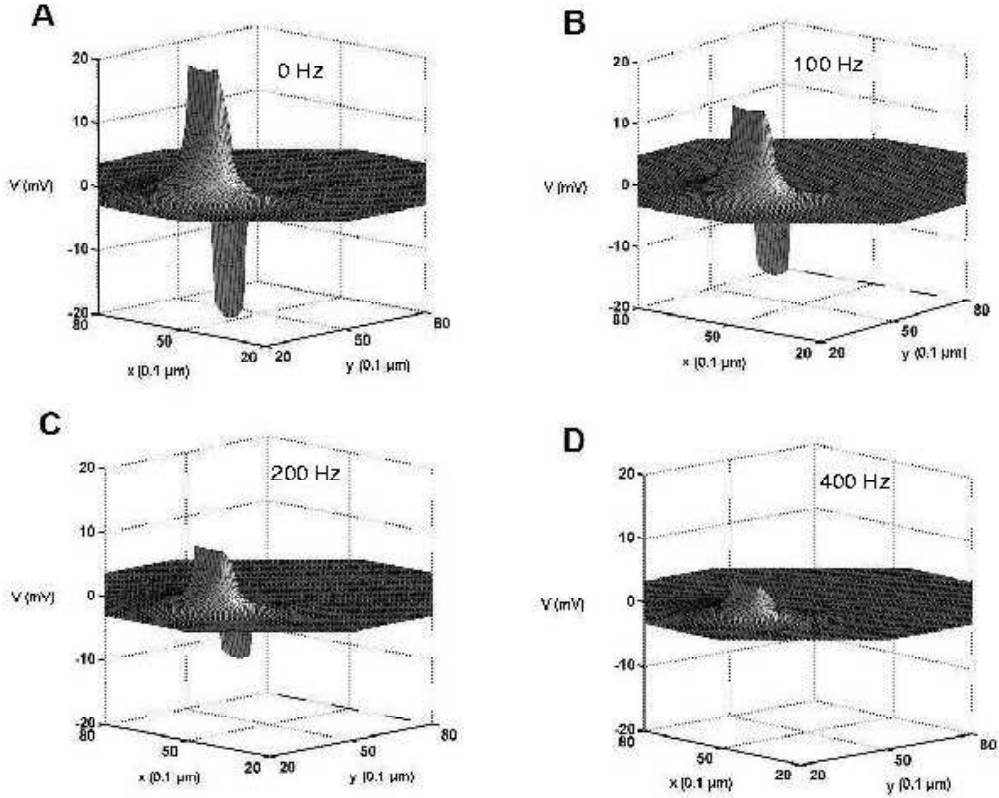}
\end{center}
\caption{Same as Fig.~[\ref{fig:obst1}], but showing the induced potential $V_{ind}$ for a time-dependent source (frequencies $f = 0, 100, 200,
400$~Hz in A,B,C,D, respectively). The induced potential tends to zero when $f$
increases.}
\label{fig:obst2}
\end{figure*}
\begin{figure*}[ht]
\begin{center}
\includegraphics[scale=0.85,angle=0]{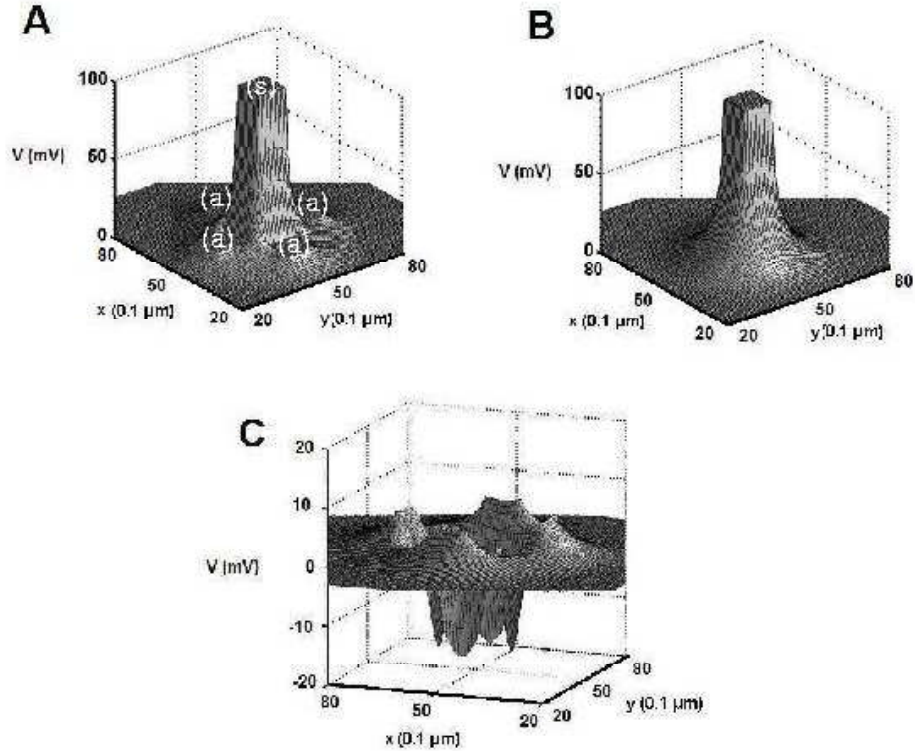}
\end{center}
\caption{Extracellular potential generated by a time-independent source
surrounded by six passive cells. Same description as in Fig.~[\ref{fig:obst1}], but for a system of a single static source (s) and 6 passive cells. 
The figure shows the potential in a horizontal plane in which 4 passive cells are visible (a). 
A: Total extracellular potential, B: Source electric potential, 
C: Induced electric potential $V_{ind}$. }
\label{fig:obst3}
\end{figure*}
\begin{figure*}[ht]
\begin{center}
\includegraphics[scale=0.750,angle=0]{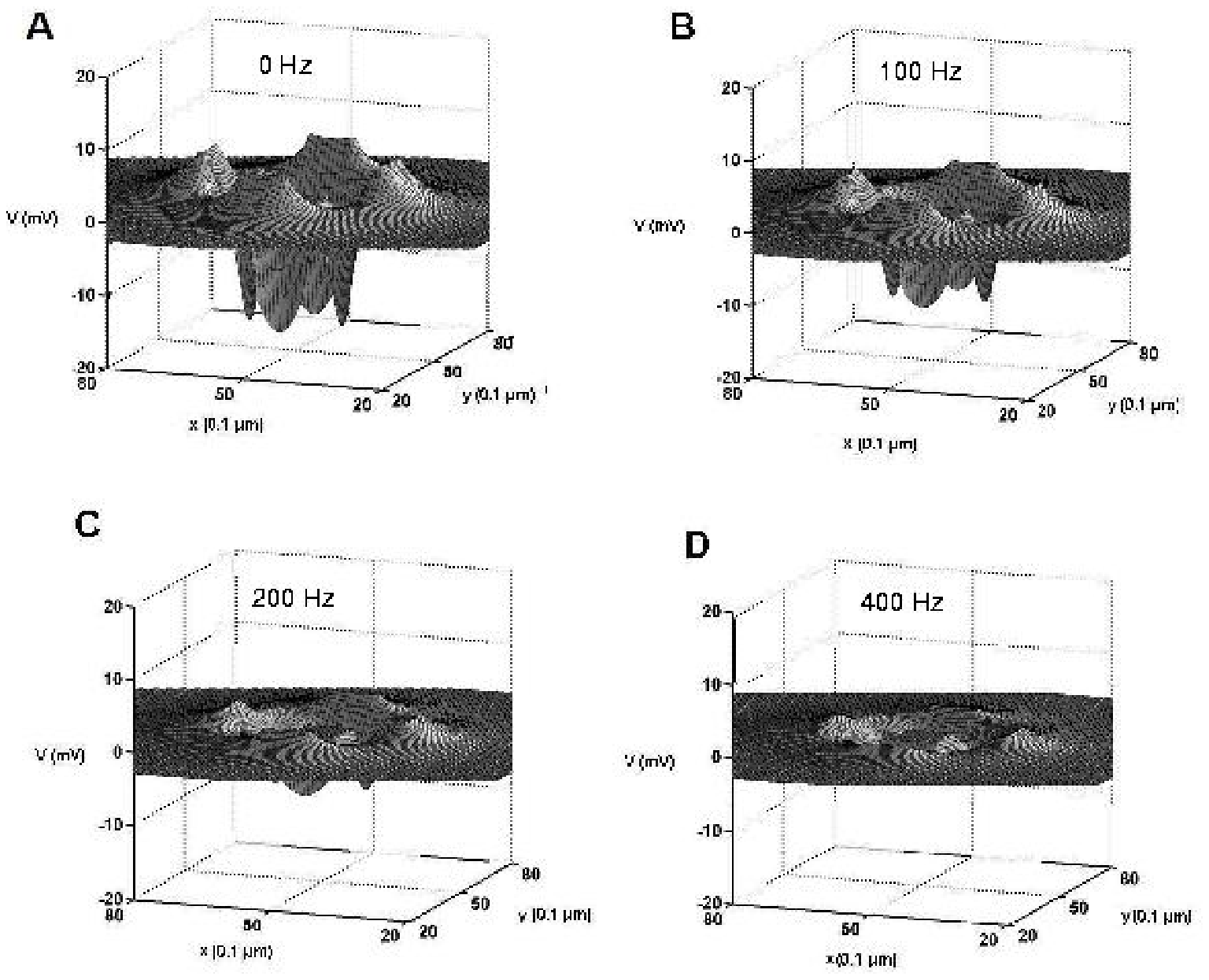}
\end{center}
\caption{Same as Fig.~[\ref{fig:obst3}], but showing the induced potential 
$V_{ind}$ for a time dependent source at different oscillation frequencies 
($f = 0, 100, 200, 400$~Hz in A,B,C,D, respectively).}
\label{fig:obst4}
\end{figure*}

\section{Attenuation as a function of distance}
\label{sec:Attenuation}

The simulations shown in the preceding section illustrated the fact that the
induced potential vanishes for very high frequencies of the source field, a fact
that can also be deduced from Eq.~[\ref{eq:DiffEqBothCases}].  In other words,
for very high frequencies ($>> T_{M}^{-1}$), the extracellular field will be
equal to the source field, since the induced field will vanish.  The space
dependence is easy to deduce in such a case, and the extracellular potential
attenuates with distance according to a $1/r$ law, as if the source was
surrounded only by conducting fluid.

However, for very low frequencies ($<< T_{M}^{-1}$), the space dependence of the
extracellular potential will be a complex function depending on both the $1/r$
attenuation of the source field, and the contribution from the induced field. 
Such a space dependence is not easy to deduce, since it depends on the spatial
arrangement of fluids and membranes around the source.  In this section, we
attempt to derive such a low-frequency space dependence for a more realistic
system of densely packed cells (illustrated in Fig.~[\ref{fig:packed}A]).  To
constrain the behavior to low frequencies, we only consider the zero-frequency
limit by using a constant source field.  We proceeded in two steps.  First, we
calculate the electric potential at the surface of a passive cell
(Section~[\ref{sec:oneSphere}]).  Second, we calculate the spatial profile of
LFPs in a system of densely packed spheres of identical shape
(Section~[\ref{sec:manySpheres}]).  

\subsection{Electric potential at the surface of passive membranes at
equilibrium}
\label{sec:oneSphere}

Let us assume a spherical passive cell embedded in a perfect dielectric medium,
and exposed to a constant electric field.  At equilibrium, we have seen above
that the effect of the electric field is to polarize the charge distribution at
the surface of the cell, such as to create a secondary electric field (see
Fig.~[\ref{fig:charge}B]), but the induced electric field is zero inside the
cell.  In this case, the conservation of charges on the surface implies:
\begin{equation}
\iint_{Surf}~\rho_{Surf}~dS = 0 ~ ,
\end{equation}
where $\rho_{Surf}$ is the charge density on the surface of the cell.  The resulting
electric potential is given by:
\begin{equation}
\label{A2}
V_{tot}(x,y,z) = V_{source}(x,y,z) 
+ \iint_{Surf} ~ \frac{\rho_{Surf}}{4\pi\epsilon~r} ~ dS
\end{equation}
where $V_{source}$ is the electric potential due to the source field, $V_{tot}$ is
the total resulting electric potential due to the source field and the induced
field, and $r$ is the distance from point $(x,y,z)$ to the center of the cell. 
Because at the center of the cell, $(a,b,c)$, we necessarily have $r=R$ (where R
is the cell's radius), the value of the resulting electric potential at the
center is given by:
\begin{equation}
V_{tot}(a,b,c)  =  V_{source}(a,b,c)
+ \iint_{Surf} ~ \frac{\rho_{Surf}}{4\pi\epsilon~R}~dS  
=  V_{source}(a,b,c) ~.
\label{eq:centerV}
\end{equation}
Thus, the electric potential at the surface of a spheric passive cell at
equilibrium equals the potential due to the primary field at the center of the
cell. In other words, the effect of the secondary field in this case perfectly
compensates the distance dependence of the primary field, such as the surface of
the cell becomes isopotential, as discussed above.

\subsection{Attenuation of electric potential in a system of packed spheres}
\label{sec:manySpheres}

\begin{figure*}[ht]
\begin{center}
\includegraphics[scale=0.6,angle=0]{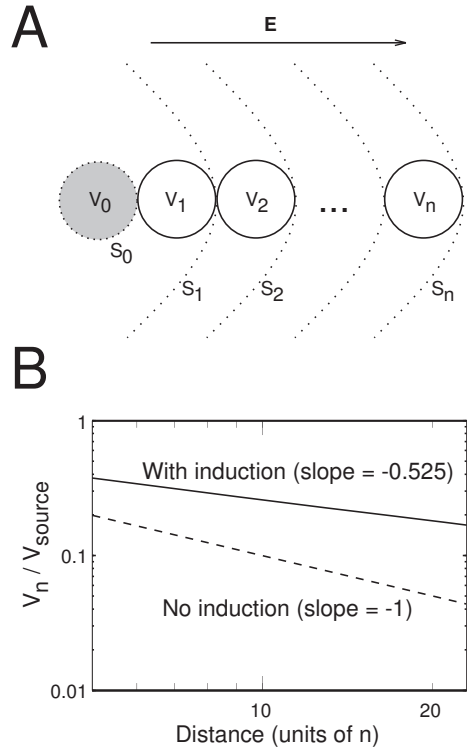}
\end{center}
\caption{Extracellular potentials as a function of distance for a system of
densely packed spherical cells. A. Scheme of the arrangement of successive
layers of identical passive cells packed around a source (in gray).  The
potential of the source is indicated by $V_0$. $V_1$, $V_2$ ... $V_n$ indicate
the potential at the surface of passive cells in layers 1,2,...$n$,
respectively.  The dotted lines indicate isopotential surfaces, which are
concentric spheres centered around the source, and which are indicated here by
$S_1$, $S_2$ ...  $S_n$. B. Extracellular potential as a function of distance
(in units of cell radius), comparing two cases: with induction (solid line,
corresponding to the arrangement schematized in A), and without induction (dashed
line; source surrounded by conductive fluid only).  Both cases predict a
different scaling of the electric potential with distance (see text for details).}
\label{fig:packed}
\end{figure*}

Keeping the assumption of a constant electric field, we now calculate how the
extracellular potential varies as a function of distance in a simplified
geometry.  We consider a system of packed spheres as indicated in
Fig.~[\ref{fig:packed}A].  From the solution of Laplace's equation, the 
extracellular potential at a distance $r$ from the source center is given by:
\begin{equation}
V(r) \ = \ \frac{k}{r}~,
\end{equation}
where $k$ is a constant, which is evaluated from the potential at the surface of
the source ($S_0$ in Fig.~[\ref{fig:packed}A]):
\begin{equation}
V(R) \ = \ V_0 \ = \ \frac{k}{R}~,
\end{equation}
where $R$ is the radius of the source.  Thus, the potential due to the source
field at a given point $r$ in extracellular space is given by:
\begin{equation}
V(r) \ = \ \frac{R \ V_0}{r}~.
\label{eq:primary}
\end{equation}

Considering the arrangement of Fig.~[\ref{fig:packed}A], if all cells of a given
layer are equidistant from the source, their surface will be at the same
potential (see Section~[\ref{sec:oneSphere}]), which we approximate as a series of
isopotential concentric surfaces ($S_1$, $S_2$ ... $S_n$ in
Fig.~[\ref{fig:packed}A]). A given layer ($n$) of isopotential cells is therefore
approximated by a new spherical source of radius $r_n$, which will polarize cells
in the following layer ($n+1$).  According to such a scheme, the potential in
layer $n+1$ is given by:
\begin{equation}
V_{n+1} \ = \ \frac{r_{n} \ V_{n}}{d_{n+1}} ~ ,
\end{equation}
where $d_{n+1}$ is the distance from the center of cells in layer $(n+1)$ to the
center of the source.  According to the scheme of Fig.~[\ref{fig:packed}A], we have
$r_n = (2n+1)R$ and $d_n = 2 n R$. Thus, we can write the following recurrence
relation:
\begin{equation}
V_{n+1} \ = \ \frac{2n+1}{2n+2} \ V_{n} ~ .
\end{equation}
Consequently:
\begin{equation}
V_{n+1} \ = \ (\prod_{j=1}^{n}~\frac{2j+1}{2j+2}) \ V_0 ~ ,
\end{equation}
which can be written, for large $n$:
\begin{equation}
V_{n} \ = \ \frac{(2n+1)!}{2^{2n} (n+1)! n!} \ V_0 
\approx \frac{2 (2n)!}{2^{2n} (n!)^{2}} \ V_0 ~ .
\end{equation}
Using Stirling's approximation, 
$n! \simeq (n/e)^{n} \sqrt{2\pi n}$ for large $n$, leads to:
\begin{equation}
V_{n} \ \simeq \ \frac{2}{\sqrt{\pi n}} ~ V_0 ~ .
\end{equation}
Thus, in a system of densely packed spherical cells, the extracellular potential
falls-off like $1/\sqrt{r}$ (Fig.~[\ref{fig:packed}B], continuous line).

In contrast, in the absence of passive cells in extracellular space, the 
electric potential is given by the source field only (Eq.~[\ref{eq:primary}]), 
which, using the same distance notations as above, is given by:
\begin{equation}
V_{n+1} \ = \ \frac{V_0}{2(n+1)} ~ .
\end{equation}
Such $1/r$ behavior is illustrated in Fig.~[\ref{fig:packed}B] (dashed line). 
Note that other theories predict a steeper decay. For instance the Debye-H\"uckel theory
of ionic solutions~\cite{Debye23} predicts a fall-off as $\exp(-k r)/r$. 

Thus, for this particular configuration, there is an important difference in the
attenuation of extracellular potential with distance.  The extracellular
potential in a system of densely packed spheres falls-off approximately like
1/$\sqrt{r}$, in contrast to a 1/$r$ behavior in a homogeneous extracellular
fluid.  Note that a 1/$\sqrt{r}$ behavior can also be found in a system in which
the source is defined as a current.  For a constant current source $I_0$, with
variations of conductivity following a spherical symmetry around the source, the
extracellular potential is given by~\cite{Bedard04}:
\begin{equation}
V(r) \ = \ \frac{I_0}{4\pi} \ \int_{r}^{\infty}\frac{1}{r^2\sigma(r) }dr ~ ,
\end{equation}
where $\sigma(r)$ is the radial profile of conductivity around the source. 
Assuming that $\sigma(r)= \sigma_0 / \sqrt{r}$, gives
\begin{equation}
V(r) \ = \ \frac{\sigma_0 I_0}{4\pi \sqrt{r}}
     \ = \ \sqrt{\frac{R}{r}} \ V_0 ~ .
\end{equation}

Consequently, the 1/$\sqrt{r}$ behavior found above is functionally equivalent to
a medium with conductivity varying like 1/$\sqrt{r}$.  This ``effective
conductivity'' is similar to that introduced in a previous study~\cite{Bedard04}.

\section{Discussion}
\label{sec:Discussion}

 From theoretical considerations and numerical simulations we have obtained the
following main results: (i) We explored the assumption that neuronal current
sources polarize neighboring cells.  This polarization produces an induced
electric field, which adds to the electric field directly produced by the
sources.  This induced field is non-negligible for biologically realistic
parameters. (ii) This induced electric field has strong frequency dependent
properties. This system is equivalent to an equivalent RC circuit and always has
low-pass filtering properties. (iii) The cut-off frequency of this low-pass
filter is determined by Maxwell's relaxation time of the membrane surfaces
surrounding neuronal sources.  (iv) Consequently, the attenuation of high
frequencies will not be influenced by this induced field, and will be the same as
if neurons were embedded in a homogeneous extracellular medium. On the other
hand, the attenuation of low frequencies will depend on the induced field, and
will therefore depend on the geometry of fluids and membranes in the
extracellular space.

In previous work, we showed that inhomogeneities of conductivity or permittivity
in extracellular space could give rise to strong frequency filtering
properties~\cite{Bedard04}.  However, low-pass or high-pass filters could be
obtained depending on the profiles of conductivity used in this model.  In the
present model, a low-pass filter is predicted, consistent with experiments (which
never evidenced a high-pass filter).  However, the fact that frequency filtering
properties are due to the alternance of high-conductive fluids and low-conductive
membranes in the present model is compatible with the inhomogeneity of
conductivity postulated in the previous model.  One main difference with the
previous model is that here, we explicitly considered a non-zero charge density
on neighboring membranes.

Another similarity with the previous model concerns the mechanism of attenuation
of LFPs with distance.  The primary or source field experiences a steep
attenuation for all frequencies, and the electric potential will attenuate with
distance $r$ following a $1/r$ law, similar to the attenuation in a homogeneous
conducting fluid~\cite{Nunez81}.  On the other hand, the induced field will be
subject to a different law of attenuation with distance, which will depend on the
spatial arrangement of passive cells around the source.  In the previous model it
depended on the particular conductivity profile~\cite{Bedard04}. In general, for
densely packed cellular membranes around the source, one will have a law of
attenuation which will be of $1/r^{\alpha}$, where $\alpha<1$ may depend on the
frequency $\omega$.  For example, we found $\alpha \sim 0.5$ for spherical cells
in Fig.~[\ref{fig:packed}]) for low $\omega$.  Thus, both models predict the
following scenario: high frequencies follow an attenuation in $1/r$, similar to a
homogeneous extracellular fluid, but low frequencies follow a slower attenuation
profile, because these frequencies are ``transported'' by the induced field.  

It should be noted that this model represents a strong approximation of the
actual complexity of the mechanisms involved in LFP generation.  A first
approximation was to consider passive cells as spheres, neglecting their
morphological complexity.  Another assumption was that the electric field results
only from neuronal current sources, while other cell types, such as glial cells,
also contains ion channels and may influence LFPs.  A third approximation was to
neglect the effect of variations of extracellular ionic concentrations (like
potassium buffering by glial cells), which may also influence LFP activity,
especially at low frequencies (in addition to local variations of conductivity). 
It is difficult to estimate the consequences of these different approximations,
but the polarization effects described here should occur in more complex
situations, so it is likely that the present considerations apply to more complex
geometries and current sources.  This should be tested by numerical simulations,
for example by constraining the model with 3D reconstructions of extracellular
space from electron microscopic measurements.

The scenario outlined above of a ``transport'' of low frequencies can be tested
experimentally in different ways.  First, measuring the decay of specific
frequency components of LFPs with distance, and in particular how they differ
from the $1/r$ law, should yield direct information on how much extracellular
space deviates from a homogeneous conducting fluid.  The attenuation profile
should also depend on the spatial arrangement of successive fluids and membranes
in the extracellular fluid.  It is conceivable to inject sinusoidal currents (of
low amplitude to allow induction effects) of different frequencies using a
microelectrode and measure the LFP produced at increasing distances, to obtain
the law of attenuation as a function of frequency (i.e., $\alpha(\omega)$).  The
comparison with model predictions should tell us to what extent the model
predicts the correct attenuation.

The equivalent RC circuit investigated above also deserves some comment. It is
well known that an electric circuit consisting of a resistor and a capacitor,
which stores electric energy, introduces a phase difference between current and
potential, and has characteristic low-pass filtering properties.  The fact that
such a simple RC circuit may be used as a model of frequency filtering of
cerebral tissue is, however, more surprising.  In fact, if we construct an
arrangement of densely packed passive cells around a current source (such as in
Fig.~[\ref{fig:packed}A]), the system will be equivalent to a series of RC
circuits, each representing one ``layer'' of cells.  In this context, it may be
interesting to relate the type of spatial arrangement with the type of equivalent
circuit(s) obtained, for example by considering different cases such as cells or
random diameters, various shapes, or even by using 3-D morphological data from
real brains obtained by serial reconstructions from electron microscopy.  This
type of investigation constitutes a possible extension of the present work.  In
addition, we have considered a source being given by a single neuron. In reality,
the source of LFP activity is a system of many neurons with complicated phase
relationships between their activities. An extension of our model in this
direction would also be a most important step to take.

A second model prediction that should be testable experimentally concerns the
predicted cut-off frequency.  This information can be related to experimentally
recorded LFPs which could be analyzed using Fourier analysis to yield information
about the cut-off frequency.  However, such an analysis should be done by
comparing different network states, to make sure that the cutoff frequency is not
dependent on activity but rather depends on structural parameters.  If correctly
done, this analysis should provide an estimate of the Maxwell's relaxation time
for the surface of membranes, or equivalently the tangential conductivity of
membrane, for which there exists presently no experimental measurement.  Work in
this direction is under way.  

Finally, we would like to conclude by suggesting that these theoretical
predictions in conjunction with measurements of frequency-dependent field
potentials may lead to new ways for detecting anomalies, for example due to
degenerative diseases like Alzheimer or brain tumors.  If such diseases cause
structural changes in brain tissue, e.g.\ creation of vacuoles, this will
manifest itself by changes of electric properties like the attenuation of
low-frequency components with distance.  These changes should be visible in
measurements of EEG, ECoG or LFP signals from multielectrodes in the area of
affected brain tissue.  Detecting such changes supposes a prior understanding of
the spectral filtering properties of these electric signals, which is one of the
main motivation of the present work.

\section{Appendix}
\label{sec:Appendix}
Here we discuss the Heaviside method as well as the Fourier transformation for 
the purpose to determine the asymptotic behavior of the solution of a class of 
differential equations.
We want to determine the asymptotic behavior of the following differential 
equation,
\begin{equation}
\label{eq:AppC-1}
\frac{\partial V(x,t)}{\partial t} + \frac{\sigma}{\epsilon} V(x,t) 
= \frac{\sigma}{\epsilon} B(x) H(t) \exp(i \omega t) ~ .
\end{equation}
The Heaviside function $H(t)$ or $\theta$-function $\theta(t)$ is defined as
\begin{equation}
\label{eq:AppC-2}
H(t) = 
\left\{ 
\begin{array}{c}
0 ~ \mbox{if} ~ t \leq 0 \\ 1 ~ \mbox{if} ~ t > 0
\end{array}
\right.
\end{equation}
Let us start out by considering the following differential equation,
\begin{equation}
\label{eq:AppC-3}
\frac{\partial V(x,t)}{\partial t} + \frac{\sigma}{\epsilon} V(x,t) 
= \frac{\sigma}{\epsilon} B(x) H(t) ~ .
\end{equation}
One verifies that the solution for $t > 0$ has the form
\begin{equation}
\label{eq:AppC-4}
V(x,t) = B(x) [ 1 - \exp( - \frac{\sigma}{\epsilon} t) ]  ~ .
\end{equation}
For $t \to \infty$, one finds 
\begin{equation}
\label{eq:AppC-5}
V(x,t) \sim_{t \to \infty} B(x) ~ . 
\end{equation}
That is $B(x)$ is the boundary condition at $t=\infty$ for the solution of 
Eq.~[\ref{eq:AppC-3}].
On the other hand, if we consider the differential equation
\begin{equation}
\label{eq:AppC-6}
\frac{\partial V(x,t)}{\partial t} + \frac{\sigma}{\epsilon} V(x,t) = 0  ~ ,
\end{equation}
and impose the boundary condition
\begin{equation}
\label{eq:AppC-7}
V(x,t=0) = B(x) ~ ,
\end{equation}
one obtains the solution
\begin{equation}
\label{eq:AppC-8}
V(x,t) = B(x) \exp(-\frac{\sigma}{\epsilon} t) ~ .
\end{equation}
Now let us consider as inhomogeneous term an oscillating step function,
\begin{equation}
H(\sin(\omega t)) ~ .
\end{equation}
This function hops between zero and 1 with a frequency $\omega$ thus giving a
oscillating step function. A function of this kind (however shifted and
stretched) is shown in Fig.~[\ref{fig:squarewave}A]. Now let us consider
Eq.~[\ref{eq:AppC-3}] with solution given by Eq.~[\ref{eq:AppC-4}]. The
inhomogeneous term of the differential equation (r.h.s.) is constant for positive
time. The solution increases with time to approach an asymptotic value.  Next
consider  Eq.~[\ref{eq:AppC-6}] with solution given by Eq.~[\ref{eq:AppC-8}]. The
inhomogeneous term of the differential equation (r.h.s.) is zero for positive
time. The solution decreases with time to approach the asymptotic value zero. 
Thus, when we consider now the differential equation
\begin{equation}
\label{eq:AppC-9}
\frac{\partial V(x,t)}{\partial t} + \frac{\sigma}{\epsilon} V(x,t) 
= \frac{\sigma}{\epsilon} B(x) H(t) H(\sin(\omega t)) ~ ,
\end{equation}
we see that its inhomogeneous term is an oscillating  step function. That means
its inhomogeneous term alternates between the inhomogeneous term of
Eq.~[\ref{eq:AppC-3}] and that of Eq.~[\ref{eq:AppC-6}]. Consequently, the solution
of Eq.~[\ref{eq:AppC-9}] alternates between that given by Eq.~[\ref{eq:AppC-4}] and
that by Eq.~[\ref{eq:AppC-8}], i.e. its increases and then decreases
exponentially. Such a solution is shown in Fig.~[\ref{fig:squarewave}B].

\noindent Now let us consider the differential equation [\ref{eq:AppC-1}]. We want 
to compute its asymptotic behavior for large positive time. 
We consider the homogeneous equation
\begin{equation}
\label{eq:AppC-10}
\frac{\partial u(x,t,t_0)}{\partial t} 
+ \frac{\sigma}{\epsilon} V(x,t,t_0) = 0  ~ .
\end{equation}
It has a solution
\begin{equation}
\label{eq:AppC-11}
u(x,t,t_0) = B(x) \exp(i \omega t_0 - \frac{\sigma}{\epsilon} t) ~ ,
\end{equation}
which satisfies
\begin{equation}
\label{eq:AppC-12}
u(x,t=0,t_0) = B(x) \exp(i \omega t_0 ) ~ .
\end{equation}
Based on the solution $u(x,t,t_0)$, Eq.~[\ref{eq:AppC-11}], of the homogeneous 
Eq.~[\ref{eq:AppC-10}], one can express the solution of the inhomogeneous 
Eq.~[\ref{eq:AppC-1}] by
\begin{eqnarray}
\label{eq:AppC-13}
V(x,t) &=& \int_{0}^{t} d t_0 ~ B(x) \exp\left( i \omega t_0 
- \frac{\sigma}{\epsilon} (t - t_0) \right)
\nonumber \\
&=& \exp( i \omega t ) ~ 
\int_{0}^{t} d \tau ~ B(x) \exp( - i \omega \tau 
- \frac{\sigma}{\epsilon} \tau ) ~ .
\end{eqnarray}
Its asymptotic behavior is given by 
\begin{eqnarray}
\label{eq:AppC-14}
V(x,t) 
&\sim_{t \to \infty}& 
\exp( i \omega t ) ~ \int_{0}^{t} d \tau ~ 
B(x) \exp( - i \omega \tau - \frac{\sigma}{\epsilon} \tau ) 
\nonumber \\
&\sim_{t \to \infty}&   
\frac{1}{i \omega + \frac{\sigma}{\epsilon} } ~ B(x) ~ \exp(i \omega t) ~ .
\end{eqnarray}
Then from the definition of the transfer function,
Eq.~[\ref{eq:DefTransferFct}], 
one obtains
\begin{equation}
\label{eq:AppC-15}
F_{TM}(\omega) = \frac{1}{1 + i \omega \epsilon /\sigma} ~ .
\end{equation}

\bigskip

\subsection*{Acknowledgments}
H.K. is grateful for support by NSERC Canada. A.D. has been supported by CNRS
and HFSP. We acknowledge discussions with Yves DeKoninck and Arnaud Delorme.


\end{document}